\documentstyle[12pt,epsfig,axodraw,a4]{article} 
\def\bild#1#2{    
        \vspace*{-5mm}
        \begin{center}
        \begin{math}
        \epsfxsize#2cm
        \epsffile{#1}
        \end{math}
        \end{center}  }
\newcommand{\vs}{\vspace{-0.25cm}}
\begin{document} 
\begin{center}
\large{\bf Induced pseudoscalar form factor of the nucleon at two-loop order in
chiral perturbation theory}

\bigskip

N. Kaiser\\

\medskip

{\small Physik Department T39, Technische Universit\"{a}t M\"{u}nchen,
    D-85747 Garching, Germany}

\end{center}

\medskip

\begin{abstract}
We calculate the imaginary part of the induced pseudoscalar form factor of the
nucleon $G_P(t)$ in the framework of two-loop heavy baryon chiral perturbation 
theory. The effect of the calculated three-pion continuum on the pseudoscalar 
constant $g_P = (m_\mu/2M) G_P(t=-0.877m_\mu^2)$ measurable in ordinary muon 
capture $\mu^-p\to \nu_\mu n$ turns out to be negligibly small. Possible 
contributions from counterterms at two-loop order are numerically smaller than
the uncertainty of the dominant pion-pole term proportional to the 
pion-nucleon coupling constant $g_{\pi N}= 13.2\pm 0.2$. We conclude that a 
sufficiently accurate representation of the induced pseudoscalar form factor of
the nucleon at low momentum transfers $t$ is given by the sum of the pion-pole
term and the Adler-Dothan-Wolfenstein term: $G_P(t) = 4g_{\pi N} M f_\pi/
(m_\pi^2 -t)- 2g_A M^2 \langle r_A^2 \rangle/3$, with $\langle r_A^2\rangle =
(0.44 \pm 0.02)$\,fm$^2$ the axial mean  square radius of the nucleon.   
\end{abstract}


\bigskip

To be published in: {\it The Physical Review C (2003), Brief Report}

\bigskip 

The structure of the nucleon as probed by charged weak currents is encoded in
two form factors, the axial and the pseudoscalar ones. To be specific consider
the matrix element of the isovector axial current between nucleon states:
\begin{equation} \langle N(p+k)|\bar q \gamma^\nu \gamma_5\tau_a q|N(p)\rangle 
= \bar u(p+k) \bigg[  \gamma^\nu G_A(t) +{k^\nu \over 2M} G_P(t)\bigg] \gamma_5
\tau_a u(p) \,, \end{equation}
where $t=k^2$ denotes the Lorentz-invariant squared momentum transfer and 
$u(p)$ stands for a Dirac-spinor. $M=938.92\,$MeV is the (average) nucleon 
mass. The form in eq.(1) follows from Lorentz-covariance, isospin conservation 
and the discrete symmetries C, P and T. $G_A(t)$ is called the axial form 
factor of the nucleon and $G_P(t)$ the induced pseudoscalar form factor of the 
nucleon. While experimentally much attention has been focussed on the first
one, the latter is generally believed to be well understood in terms of 
pion-pole dominance as indicated from ordinary muon capture experiments $\mu^- 
p \to  \nu_\mu n$ (see e.g. refs.\cite{review,omc,muoncap}). The pseudoscalar
coupling constant $g_P$ as measured in ordinary muon capture is defined via:  
\begin{equation} g_P = {m_\mu \over 2M} G_P(t_\mu)\,, \qquad t_\mu = {M_n^2
m_\mu \over M_p+m_\mu }- M_p m_\mu = -0.877\,m_\mu^2 = -0.502\, m_\pi^2 \,,
\end{equation}
with $t_\mu$ the Lorentz-invariant squared momentum transfer if the proton and
muon are initially at rest. $m_\mu = 105.66\,$MeV is the muon mass, 
$M_p = 938.27\,$MeV the proton mass, $M_n = 939.57\,$MeV the neutron mass and 
$m_\pi = 139.57\,$MeV the charged pion mass. 

Chiral perturbation theory allows to calculate systematically the corrections
to the dominant pion-pole term in $G_P(t)$ (see eq.(7) below). At one-loop 
order this correction is uniquely expressed in terms of the mean square 
axial radius of the nucleon $\langle r_A^2\rangle$ by making use of the chiral 
Ward identity of QCD \cite{axialward}. Exactly the same term, derived
originally by Adler, Dothan and Wolfenstein \cite{dothan}, is also found in 
the small scale expansion of ref.\cite{sse} where additional diagrams with
intermediate $\Delta(1232)$-isobars contribute. While the one-loop prediction
for the pseudoscalar coupling constant $g_P = 8.4\pm0.2 $ \cite{axialward,sse}
is consistent with the earlier result of the Saclay experiment $g_P =8.7 \pm
1.9$ \cite{omc} a reanalysis \cite{review} of that experiment using the modern
world average of the muon mean life time gives the enhanced value $g_P=10.6 \pm
2.7$. For further details on that and the conflicting situation concerning
radiative muon capture $\mu^- p \to \nu_\mu n\gamma$ see the recent review of
Gorringe  and Fearing \cite{review} and also ref.\cite{muoncap}.

\bigskip

\bild{gpfig.epsi}{13}
{\it Fig.1: Two-loop diagrams contributing to the imaginary part of the induced
pseudoscalar form factor of the nucleon $G_P(t)$. Dashed and solid lines denote
pions and nucleons, respectively. The wiggly line symbolizes the external 
isovector axial source. The combinatoric factor of the first four diagrams is 
1/6 and the next four graphs have the combinatoric factor 1/2. The last two
diagrams scale as $g_A^3$ whereas the other eight graphs scale as $g_A$.}

\bigskip

The purpose of the present short paper is to investigate the two-loop
corrections to the induced pseudoscalar form factor $G_P(t)$ in order to 
clarify whether these could affect (numerically) the theoretical prediction for
$g_P$ used so far. The essentially new feature at two-loop order is a
nonvanishing imaginary part Im\,$G_P(t)$ for $t>9m_\pi^2$  which originates 
from the (direct and indirect) coupling of the isovector axial current to the
three-pion intermediate state. The pertinent ten topologically distinct
two-loop diagrams generated by leading order vertices of the effective chiral
Lagrangian ${\cal L}_{\pi\pi}^{(2)}+ {\cal L}_{\pi N}^{(1)}$ are shown in 
Fig.\,1. The Feynman rules for the relevant interaction vertices can be found 
in appendix A of ref.\cite{unser}.

Let us now turn to the evaluation of the imaginary part Im\,$G_P(t)$ from the
two-loop diagrams shown in Fig.\,1. Application of the Cutkosky cutting rules 
gives the spectral function Im\,$G_P(t)$ as an integral of the product of
$axial source \to 3\pi$ and $3 \pi \to \bar NN$ transition amplitude over the
Lorentz-invariant three-pion phase space. Some details about these techniques 
can be found in refs.\cite{spectral,3pi} where the same method has been used 
to calculate the (two-loop) spectral functions of the isoscalar electromagnetic
nucleon form factors and the $3\pi$-exchange nucleon-nucleon potential. The
choices $\epsilon\cdot k = 0$ and $\epsilon^\nu = k^\nu$ for the polarization
vector $\epsilon^\nu$ of the external isovector axial source allow a separation
of Im\,$G_A(t)$ and Im$[G_A(t)+t\,G_P(t)/4M^2]$. Alternatively, one can use 
projection operators and tensorial integrals over the $3\pi$-phase space can be
reduced to scalar ones (see eq.(13) in ref.\cite{spectral}). The pertinent 
three-body phase space integrals are most conveniently  performed in the 
three-pion center-of-mass frame. The corresponding on-mass-shell four-momenta 
of the three pions read in this frame: $k_1^\nu = (\omega_1,\vec k_1\,)$, 
$k_2^\nu = (\omega_2,\vec k_2\,)$ and $k_3^\nu = (\sqrt t-\omega_1-\omega_1, 
-\vec k_1-\vec k_2\,)$. The mass-shell condition $k_3^2 = m_\pi^2$ determines  
the cosine of the angle between $\vec k_1$ and $\vec k_2$ (called $z$) as:   
\begin{equation} z\, k_1 k_2 = \omega_1 \omega_2-\sqrt t( \omega_1+\omega_2)
+{1\over 2}(t+m_\pi^2) \,,  \qquad k_{1,2} = \sqrt{ \omega^2_{1,2}-m_\pi^2} \,.
\end{equation}
The ten diagrams in Fig.\,1 fall into two classes. The first eight diagrams
carrying the common prefactor $g_A/f_\pi^4$ give rise altogether to the 
following contribution to the imaginary part of the induced pseudoscalar form 
factor of the nucleon:
\begin{equation} {\rm Im\,}G_P^{(1)}(t) = {g_{\pi N} M \over (2\pi f_\pi)^3}
\int \limits_{z^2<1} \!\!\!d\omega_1d\omega_2\bigg\{{1\over 18}-{m_\pi^4 
\over 12(t-m_\pi^2)^2}+{4 \omega_1^2 -m_\pi^2 \over 6t }+{\omega_1^2(3m_\pi^2 
-t) \over (t-m_\pi^2)^2} +{2m_\pi^2\omega_1\omega_2 z k_2 \over t(t-m_\pi^2) 
k_1} \bigg\} \,. \end{equation}
Here, $f_\pi = 92.4\,$MeV denotes the pion decay constant and we have employed
the Goldberger-Treiman relation: $g_{\pi N} f_\pi = g_A M$. The inequality
$z^2<1$ determines the kinematically allowed region in the $\omega_1\omega_2
$-plane (which is bounded by a cubic curve) together with the obvious
kinematical constraints $m_\pi < \omega_{1,2} < \sqrt t-2m_\pi$ and $2m_\pi < 
\omega_1 + \omega_2 < \sqrt t-m_\pi$. Furthermore, one derives from the last 
two diagrams in Fig.\,1 which are proportional to $g_A^3/f_\pi^4$ the following
contribution to the imaginary part Im\,$G_P(t)$: 
\begin{eqnarray} {\rm Im\,}G_P^{(3)}(t) &=& {g_{\pi N} M g_A^2 \over (2\pi 
f_\pi)^3t} \int \limits_{z^2<1} \!\!\!d\omega_1d\omega_2\bigg\{(m_\pi^2-\sqrt t
\omega_1) \bigg( z+{k_2\over k_1}\bigg) {\arccos(-z) \over \sqrt{1-z^2}}  
\nonumber \\ && + {k_1^2 \over 3}+{t\over 9} +{m_\pi^2\over t-m_\pi^2}\bigg(
{7\over 8} \sqrt t-\omega_1-\omega_2 \bigg)\bigg[2\omega_1 {z k_2\over k_1} +
\sqrt t \nonumber \\ && + \Big((t+m_\pi^2)(4\omega_1-\sqrt t)
-4 \sqrt t \omega_1\omega_2\Big){ \arccos(-z) \over 2k_1k_2\sqrt{1-z^2}}
\bigg] \bigg\} \,. \end{eqnarray}
In the chiral limit $m_\pi = 0$ the total (two-loop) spectral function 
Im\,$G_P(t)$ shows a simple linear $t$-dependence of the form:
\begin{equation} {\rm Im\,}G_P(t)|_{m_\pi = 0} = -{4M^2 \over t}{\rm Im\,}G_A
(t)|_{m_\pi = 0}= {g_{\pi N} M \, t \over 9(8 \pi f_\pi)^3} \bigg[ 1-g_A^2
\bigg(1+{64\pi^2\over35}\bigg) \bigg] \simeq -{3t \over M^2} \,. \end{equation}
The first part of this equation follows from the fact that the combination 
$G_A(t)+t\,G_P(t)/4M^2$ is the form factor of the divergence of the isovector 
axial current, which vanishes in the chiral limit $m_\pi=0$ (QCD chiral Ward 
identity). The (two-loop) result for Im\,$G_A(t)|_{m_\pi = 0}$ has been taken
over from eq.(27) in ref.\cite{spectral}.  

In Fig.\,2 we show by the full line the total imaginary part Im\,$G_P(t)$ 
calculated from eqs.(4,5) after division by a factor $t$. The horizontal dashed
line in Fig.\,2 indicates the asymptotic behavior of Im\,$G_P(t)/t$ for $t\to
\infty$. 

\bigskip

\bild{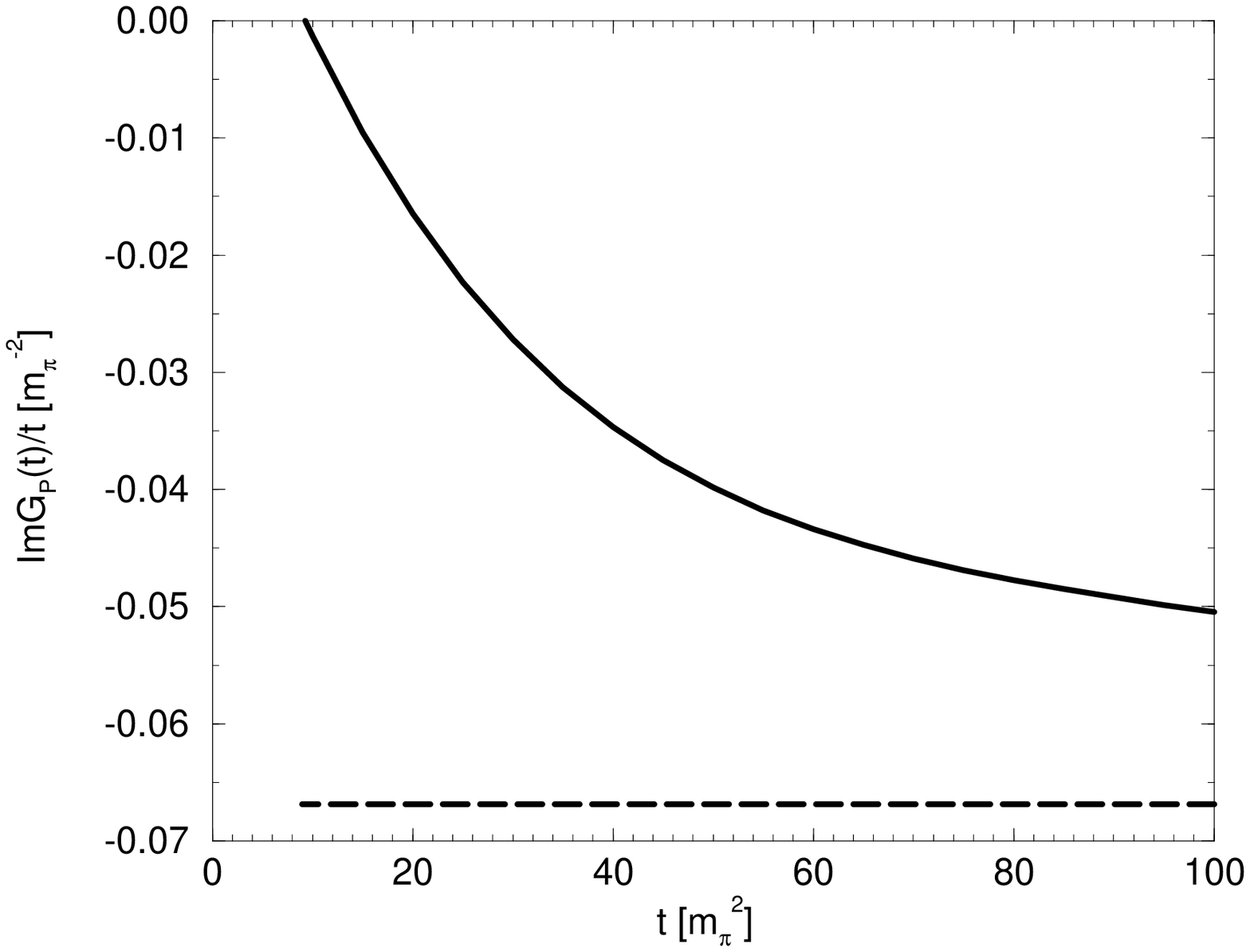}{12}
{\it Fig.2: The spectral function Im\,$G_P(t)$ of the induced pseudoscalar form
factor of the nucleon divided by $t$. The horizontal dashed line indicates its
asymptotic form obtained by taking the chiral limit $m_\pi=0$.}

\bigskip

With the help of the spectral function Im\,$G_P(t)$ the complete two-loop
representation of the induced pseudoscalar form factor of the nucleon can be
written as: 
\begin{equation} G_P(t) = {4g_{\pi N} M f_\pi \over m_\pi^2 -t} -{2\over 3} g_A
M^2 \langle r^2_A\rangle +{M^2 \over (2\pi f_\pi)^4} ( \zeta_0 m_\pi^2 +
\zeta_1 t) +{t^2 \over \pi} \int_{9m_\pi^2}^\infty dt' {{\rm Im\,}G_P(t') \over
t'^2(t'-t-i0^+)}\,. \end{equation}
The first two terms are the well-known pion-pole term and
Adler-Dothan-Wolfenstein term \cite{axialward,dothan}. The parameters $g_{\pi 
N}, f_\pi, m_\pi, \langle r^2_A\rangle$ etc. are to be understood as the
physical ones including their individual one- and two-loop chiral corrections. 
These (not explicitly calculated) two-loop renormalization effects come along  
with the real parts of the diagrammatic amplitudes. Note that the dispersion
integral in eq.(7) requires two subtractions because of the asymptotic linear
growth of the imaginary part Im\,$G_P(t)$. The third term in eq.(7) involving
the two dimensionless low-energy constants $\zeta_0$ and $\zeta_1$ subsumes all
polynomial contributions which arise from (tadpole-type) loop diagrams and 
possible chiral-invariant counterterms (beyond renormalizing the 
Adler-Dothan-Wolfenstein term). The prefactor of this term is chosen such that 
the negative mass dimension of the counterterm coupling strength is accounted 
for by appropriate powers of the chiral symmetry breaking scale $\Lambda_\chi 
= 2\sqrt{2}\pi f_\pi \simeq 0.82\,$GeV. Based on naturalness arguments one 
expects that the dimensionless low-energy constants $\zeta_{0,1}$ are of order
one. Indeed the same considerations applied to the Adler-Dothan-Wolfenstein 
term give for the analogous dimensionless low-energy constant at one-loop
order: $-(2\pi f_\pi)^2 g_A \langle r^2_A\rangle/3 = - g_A(4\pi f_\pi/M_A)^2
\simeq -1.63$. Here, we have inserted the value of the axial dipole mass $M_A =
(1.03 \pm 0.02)\,$GeV as extracted in ref.\cite{liesenfeld} from
(quasi)-elastic neutrino and antineutrino scattering experiments. 

Let us now turn to numerical results. From the twice-subtracted dispersion 
integral in eq.(7) one gets a tiny contribution to the pseudoscalar coupling 
constant of $\delta g_P \simeq -1\cdot 10^{-5}$. The extreme smallness of this
number comes partly from the proximity of the chosen subtraction point $t_0=0$
to $t_\mu = -0.502\,m_\pi^2$. Nevertheless, when varying the subtraction point
(via the substitution $t^2/t'^2 \to (t-t_0)^2/(t'-t_0)^2$ in eq.(7)) in the 
broad range $-24 m_\pi^2 <t_0 < 9m_\pi^2$ the contribution of the dispersion
integral to the pseudoscalar coupling constant stays in magnitude smaller than
one percent, $0 > \delta g_P > -1\cdot 10^{-2}$. Note that a change of the
subtraction point $t_0$ is equivalent to changes of the low-energy constants
$\zeta_{0,1}$ parameterizing the polynomial piece in eq.(7). Since we cannot
accurately determine the coefficients $\zeta_{0,1}$ we turn here the argument 
around and ask only for some upper bound. For example, in order to cause a
correction of $\delta g_P= 0.09$, corresponding to a relative $1\%$ change of
$g_{\pi N}$ in the pion-pole term, the relation $2\zeta_0- \zeta_1 \simeq 21$
must hold. Such values of $\zeta_{0,1}$ exceed the expectation from naturalness
already by one order of magnitude. 

The delta-nucleon mass-splitting $\Delta=293\,$MeV introduces another small 
scale to the problem. The systematic power scheme inherent to the small scale 
expansion of refs.\cite{muoncap,sse} ensures however that the contribution to 
$G_P(t)$ from two-loop diagrams with intermediate $\Delta(1232)$-excitations is
a homogeneous function of degree one in the three variables $(t, m_\pi^2, 
\Delta^2)$. Consequently, possible negative powers of $\Delta$ get always 
overcompensated by two more powers of the numerically smaller scales $m_\pi$ 
and/or $\sqrt{|t_\mu|}$ in $G_P(t_\mu)$. 

One may therefore conclude that all two-loop corrections to the pseudoscalar
coupling constant $g_P$ are numerically insignificant. A sufficiently accurate
representation of $G_P(t)$ at low momentum transfers $t$ is given by the sum of
the pion-pole term and the Adler-Dothan-Wolfenstein term. Using for the $\pi
N$-coupling constant $g_{\pi N} =13.2\pm 0.2$ \cite{ericson,pavan} which is
consistent with recent results from $\pi N$-dispersion relation analyses
\cite{pavan} and $\langle r^2_A \rangle = 12/M_A^2 =(0.44 \pm 0.02)$\,fm$^2$
\cite{liesenfeld} for the axial mean square radius one gets in this case:     
\begin{equation} g_P = 8.3 \pm0.2\,. \end{equation}
The major theoretical uncertainty of $g_P$ comes obviously from the 
$\pi N$-coupling constant $g_{\pi N}$ entering the dominant pion-pole term.

Let us finally consider the form factor of the nucleon related to the 
divergence of the  isovector axial current: 
\begin{equation} G_A(t)+{t\over 4M^2} \,G_P(t) = {g_A m_\pi^2 \over m_\pi^2 -t}
\, D(t) \,. \end{equation}
The prefactor on the right hand side expresses the vanishing of this form
factor in the chiral limit $m_\pi = 0$ as well as the presence of a pion-pole 
contribution. The imaginary part Im\,$D(t)$ (at two-loop order) can be easily
constructed from the expressions for Im\,$G_P(t)$ given here in eqs.(4,5) as 
well as from the formula for Im\,$G_A(t)$ written in eq.(26) of 
ref.\cite{spectral}. As a further non-trivial result we give here only 
the spectral function Im\,$D(t)$ in the chiral limit:
\begin{equation}{\rm Im}\,D(t)|_{m_\pi = 0} = {2 t^2 \over 9\pi^3 (8 f_\pi)^4 }
\bigg[ 1 +g_A^2 \bigg( 5+{68\pi^2 \over 35} \bigg) \bigg] \simeq {t^2 \over g_A
M^4} \,, \end{equation}
which may be useful for some quick order of magnitude estimates. 

In summary we have calculated in this work the imaginary part of the induced 
pseudoscalar form factor of the nucleon Im\,$G_P(t)$ at two-loop order in heavy
baryon chiral perturbation theory. Two-loop corrections to the pseudoscalar
coupling constant $g_P$ measurable in ordinary muon capture $\mu^- p \to 
\nu_\mu n$ are numerically unimportant in view of the present uncertainty of
the pion-nucleon coupling constant $g_{\pi N}$.

\bigskip
   
I thank T. Hemmert for useful discussions.

\end{document}